# Does the geographic proximity effect on knowledge spillovers vary across research fields?[1]

Giovanni Abramo
*Laboratory for Studies in Research Evaluation*
*at the Institute for System Analysis and Computer Science (IASI-CNR)*
*National Research Council of Italy*
ADDRESS: Istituto di Analisi dei Sistemi e Informatica, Consiglio Nazionale delle Ricerche,
Via dei Taurini 19, 00185 Roma - ITALY
giovanni.abramo@uniroma2.it

Ciriaco Andrea D'Angelo
*University of Rome "Tor Vergata" - Italy and*
*Laboratory for Studies in Research Evaluation (IASI-CNR)*
ADDRESS: Dipartimento di Ingegneria dell'Impresa, Università degli Studi di Roma
"Tor Vergata", Via del Politecnico 1, 00133 Roma - ITALY
dangelo@dii.uniroma2.it

Flavia Di Costa
*Research Value s.r.l.*
ADDRESS: Research Value, Via Michelangelo Tilli 39, 00156 Roma- ITALY
flavia.dicosta@gmail.com

**Abstract**
Policy makers are interested in the influence of geographic distance on knowledge flows, however these can be expected to vary across research fields. The effects of geographic distance on flows are analyzed by means of citations to scientific literature. The field of observation consists of the 2010-2012 Italian publications and relevant citations up to the close of 2017. The geographic proximity effect is analyzed at national, continental, and intercontinental level in 244 fields, and results as evident at national level and in some cases at continental level, but not at intercontinental level. For flows between Italian municipalities, citations decrease with distance in all fields. At continental level, four fields are identified having knowledge flows that grow with distance; at intercontinental level, this occurs in 26 fields. The influence of distance is more limited in the fields of Humanities and Social sciences, much more significant in the Sciences, mainly in the Natural sciences.

---

# 1. Introduction

Given the context of the "knowledge economy", policy makers have expressed increasing interest in influencing the intensity of production of new knowledge, and potentially, the speed and breadth of its diffusion. For this reason, researchers have focused on the ways in which knowledge flows are achieved, the intensity with which they manifest themselves and the factors that condition them. One of the issues for specific attention is the importance of the geographic factor in the creation and dissemination of knowledge. The geographic barriers bounding knowledge flows (Audretsch & Lehmann, 2005; Audretsch & Keilbach, 2007) have been lowered by the rapid development of information technologies, contributing significantly to savings in the costs and time of knowledge diffusion (Ding, Levin, Stephan, & Winkler, 2010).

Nevertheless, geographic distance still seems to be relevant in determining the structure of knowledge flows between territories. In their seminal article, Jaffe, Henderson and Trajtenberg (1993) found that citations to domestic patents are more likely to be domestic, and more likely to come from the same state and municipality. Twenty years later, Belenzon and Schankerman (2013) confirmed these early results, using patent citations both to university patents and scientific publications.

In shifting from the flows associated with patent citations to those of knowledge encoded in the scientific literature, one could expect the geographic factor to be less important: citations between articles, since they refer to the public content of research findings, would in theory be "placeless", i.e. not influenced by the geographic location of the cited and citing authors (Livingstone, 2003). However, a series of empirical studies have demonstrated that there a geographic proximity effect also exists in citations of scientific literature (Matthiessen, Schwarz, & Find, 2002; Börner, Penurnarthy, Meiss, & Ke, 2006; Ahlgren, Persson, & Tijssen, 2013). Pan, Kaski, and Fortunato (2012) showed that the citation flows between cities, as well as the collaboration strengths, decrease with the relative distances, following a gravity law.

Differently from most previous studies on the topic, which were mainly limited to one or few research fields, Abramo, D'Angelo and Di Costa (2019) analyzed the influence of geographic distance on knowledge flows related to the entire Italian scientific production in the period 2010-2012. Applying a gravity model, estimated using ordinary least squares (OLS), the authors show that geographic distance is an influential factor in the processes of knowledge flows between regions of the same country, and that is not negligible in "continental" flows (from Italy to European countries), but irrelevant in intercontinental flows (from Italy to non-European countries).

Frenken, Hardeman, and Hoekman (2009) offer a relevant methodological contribution, proposing an analytical framework able to distinguish between physical and other forms of "proximity", for instance "social proximity", as determinants of scientific interaction. When controlling for such forms of proximity, physical distance seems to be reduced in importance. Yan & Sugimoto (2011) observed that the steady introduction of online databases has weakened the effect of the physical distance, so that citations are now more closely dependant on the intensity of collaboration. Recently, Wuestman, Hoekman, and Frenken (2019) claimed that self-citations are an important driver of "geographic bias". Moreover, once "cognitive relatedness" (measured by the number of references shared by two publications) is accounted for, the effect of distance between citing and cited publications is weak. The authors warn about the generalizability of their findings due to the sector and time specific nature of their analysis. Also, Head, Li, and Minondo (2019) conclude that the negative impact of geographic distance on citations is "mediated" by "social relatedness". They studied how geographic distance and social ties (co-authorship, past collocation, and relationships mediated by advisors and the alma mater) affect citation patterns in mathematics, observing that when controlling for ties, the negative impact of geographic distance on citations is generally halved. The authors hypothesize that spatial proximity facilitates the creation of interpersonal links that in turn favor knowledge flows.

The contributions proposed so far in the literature are based on observations of individual fields, or at an aggregate level without distinction between fields. Therefore, we do not know if and to what extent the geographic proximity effect varies across fields. With the current work, our intention is to address this gap.

A number of reasons may explain why the role of geographic proximity might differ across research fields. Citation behavior of authors differs across research fields (Hurt, 1987; Vieira, E.S., & Gomes, 2010). Field-focused research organizations may be more or less geographically clustered, and large and numerous within clusters. Therefore, to the extent that places concentrate their research efforts on certain topics (Boschma, Heimeriks, & Balland, 2014), citations reflecting intellectual recognition will also be more geographically concentrated (Head, Li, and Minondo, 2019). Hence, the geographical proximity effect in citations may, in principle, be fully explained by the geographical concentration of intellectually related knowledge. Furthermore, in certain fields research topics might be more territory specific, addressing local needs, and therefore with more localized spillovers. Finally, when included in the analysis, self-citations amplify the role of geographic proximity (Aksnes, 2003), and it is known that self-citation rates vary across fields (Ioannidis, Baas, Klavans, & Boyack, 2019).

To conduct our investigation, we analyze the world publications citing up to the close of 2017, the Italian publications indexed in the Clarivate Analytics Italian national citation report (I-NCR), extracted from Web of Science (WoS) core collection in the period 2010-2012. For each publication (citing and cited) we associate a prevalent territory of production, as well as the WoS subject category (SC) of affiliation. The analysis of the effect of geographic distance on citation flows is carried out using gravity models estimated using OLS, for each SC (244 in all) and geographic context (national, continental and intercontinental).

The work is structured as follows. In the next section we present the data and methods of analysis; section 3 provides the results from the elaborations; section 4 closes the work with a synthesis of main results and authors' considerations on the implications of the study.

**2. Methods**

To test the influence of geographic distance on knowledge flows at SC and area level, we apply a gravitational model similar to that used by Ponds, Van Oort, and Frenken (2007) for the study of scientific collaborations between different types of institutions. The model is based on two assumptions:
- The flow of knowledge between any two territories can be measured through the citations made in the scientific production by the research centres in the first territory, to the scientific production by the research centres in the second (i.e. citations in the scientific literature of the "citing territory" to the scientific literature of the "cited territory").
- Citations between two territories increase with the amount of scientific production of both, and decrease with the distance between them.

We assign publications (cited or citing) to a territory following the criteria conceived by Abramo, D'Angelo, and Di Costa (2019), to which we refer the reader for a thorough discussion:
- For cited publications, we define a publication as "made in" a territory if the majority of its co-authors are affiliated to organizations located in that territory.
- Differently from the cited publications, for the citing publications the I-NCR reports only the address list without the link to authors. We define then a publication as "made in" a territory if the majority of its addresses refers to that territory. [2]
- Publications with no prevalent territory are excluded from the analysis.

The analysis of knowledge flows will be carried out in three distinct geographic contexts:
- the national one, in which the citing publications assigned to "Italy" are attributed to one and only one LAU (municipality)[3] of the Italian territory, always on the basis of the prevalence criterion;
- the international one, where the citing publications will be attributed to one and only one country on the basis of the prevalent NUTS0 code;[4] we will distinguish also between the continental (Europe) and the extra-continental (extra-Europe) context.

We then measure the "distances" of the citation flows, along the geodetic line[5] that joins the prevalent Italian LAU of production of the aforementioned publication with:
- the citing Italian LAU, for national analysis,
- the capital of the citing country, for international analysis.

In this work, we control for the cognitive proximity of the citing-cited publications. Previous studies measured the mass of the citing territory by the total number of publications made in that territory (Pan, Kaski, & Fortunato, 2012) or by the number of solely publications falling in the same field as the cited publication (Abramo, D'Angelo, & Di Costa, 2019). Here, we adopt a different method to measure the mass of the citing territory.

As usual, we measure the mass of the territory of the cited publication by the total number of publications of that territory falling in the same WoS subject category (SC).

Much more complex is the way we measure the mass of the territory of the citing publications. Citing publications may fall or not fall in the same SC as the cited publication. We first calculate the SC frequency distribution of all world publications citing all Italian publications within a certain SC. The mass of the territory of each citing publication is the weighted sum of that territory's publications falling in the above identified SCs, whereby the weights correspond to their frequency distribution.

---

[2] This convention has some obvious limits: a citing publication could be attributed to a given territory when in fact the authors from that territory did not reach a "majority" within the byline; the full counting of each of the authors' addresses distorts the result in the presence of authors with multiple affiliations; finally, the corresponding author ends up having twice as much weight as the others, for the simple fact that their affiliation appears twice in the address list. In order to evaluate the effect of such limits, we extracted a random sample of 1,000 cited publications from the dataset and, for each citing record of such publications (17,216 in all), we downloaded the author-affiliation field by means of the "Advanced Search" interface in the online WoS portal. The application of both conventions to such set of citing publications reveals that in 96.8% of cases the "made in" territory remains the same.

[3] The LAU level consists of municipalities or equivalent units in the 27 EU Member States.

[4] The NUTS classification (Nomenclature of territorial units for statistics) is a system subdividing the economic territory of the European Union into hierarchical levels.

[5] In the literature, this method of measuring geographic distance has been adopted in Maurseth and Verspagen, 2002; Broekel and Mueller, 2018; Ahlgren, Persson and Tijssen, 2013; Jiang, Zhu, Yang, Xu, & Jun, 2018. Some scholars have instead adopted the travel time between two points (Crescenzi, Nathan, and Rodríguez-Pose, 2016; Ponds, Van Oort & Frenken, 2007).

To exemplify, we consider all publications in the dataset falling in the SC Paleontology. Relevant world citing publications in the observed period fall in 93 different SCs[6]: 45% in Paleontology, 18.9% in Geosciences, multidisciplinary, 9.4% in Geology, 8.6% in Geography, physical and the remaining 18% are dispersed across the remaning 89 SCs. Let us assume that we want to measure the knowledge flows generated by the cited publications in Paleontology made in LAU Milan, to LAU Turin. The mass of Milan is measured by the 2010-2012 cited publications in Paleontology made in Milan (52 in all). The mass of Turin, instead, is measured by the weighted average of the 2010-2017 publications made in Turin and falling in the above 93 SCs (189 in all).

The gravity model adopted for the national analysis in each SC is:

$$C_{ij} = k \cdot \frac{M_i^\alpha M_j^\beta}{d_{ij}^\gamma} \qquad [1]$$

with:
$C_{ij}$ = number of citations to publications made in LAU $i$ by the publications made in LAU $j$
$k$ = constant
$M_i$ = total number of publications made in LAU $i$ in the 2010-2012 period
$M_j$ = weighted number of publications made in LAU $j$ in the 2010-2017 period
$d_{ij}$ = geodetic distance between cited LAU $i$ and citing LAU $j$

For the international analysis, the following distinctions apply:
$C_{ij}$ indicates the number of citations to publications made in LAU $i$ by the publications made in country $j$
$M_j$ refers to the prevalent country $j$
$M_j$ = weighted number of publications made in country $j$ in the 2010-2017 period
$d_{ij}$ is the distance between cited LAU $i$ and the capital of the citing country $j$

Applying a logarithmic transformation to all variables of equation [1], we obtain:

$$\ln(C_{ij}) = \ln(k) + \alpha \ln(M_i) + \beta \ln(M_j) - \gamma \ln(d_{ij}) + \varepsilon \qquad [2]$$

The coefficients of a log-log model represent the elasticity of the Y dependent variable with respect to the X independent variable. For example, for the distance variable ($d_{ij}$) an elasticity of one (γ = 1) indicates that a 1% increase in the distance is associated with a 1% decrease in citations exchanged, on average.

For the 2010-2012 triennium the I-NCR dataset contains 255,399 Italian publications, 184,177 of which had received at least one citation up to the close of 2017. 161,680 were assigned univocally to an Italian LAU,[7] and had received 3,002,835 total citations from 1,800,037 citing publications. The overall dataset was broken down by SC (244 in all, according to the WoS classification schema) of the hosting journal.[8] In turn, the SCs are grouped in OECD disciplinary areas (DAs, six in all) applying a category-to-category mapping available on the Incites-Clarivate Analytics portal.[9]

---
[6] Papers published in multi-category journals are full counted in each category.
[7] The remaining publications had no prevalent LAU, and have been assigned to none.

## 3. Empirical evidence

The following sections illustrate the results of the analysis at two levels of aggregation: i) by DA; ii) by SC. For each level, the analysis was carried out considering three geographical contexts: national, European and extra-European, depending on the location of the citing publications.

### 3.1 Disciplinary area level analysis

### 3.1.1 The national context

For the analysis of the national context, Table 1 shows the descriptive statistics of the variables used in the gravitational model[10] estimated for each DA.

In the period of observation, the mean citation flows between Italian municipalities vary greatly among the DAs considered, ranging from a minimum of 3.6 (Humanities) to a maximum of 41.5 (Natural sciences). Differences are mainly due to the peculiar characteristics of the DAs considered, such as the different intensity of publication and citation.

Focusing on the variable of interest $d_{ij}$, it can be observed that for all DAs the mean distance is always higher than the median, revealing a right skewed distribution. The mean distance ranges from a minimum of 320 km (Humanities) to 373km (Natural sciences). The maximum distance of citation flows registered between two Italian municipalities ranges between 1022 km (Humanities) and 1119 km (Medical and health sciences). To contextualize these figures, it should be observed that the maximum geographic distance between two LAUs, from extreme southern to northern Italy, is 1271 km (Lampedusa, Vipiteno).

In summary, it is clear from the data observed that compared to the other DAs, in the Social sciences and Humanities the average distances of citation flows between national organisations are significantly smaller.

---

[8] Publications in multi-category journals are assigned to each category.
[9] http://help.prod-incites.com/inCites2Live/5305-TRS.html, last access 22 January 2020.
[10] The results of the analysis in the European and extra-European contexts are presented in appendix Tables A1 and A2.

*Table 1: Descriptive statistics for the variables of the gravitational model applied to the national context, by disciplinary area*

| Area | var | Obs | mean | p25 | p50 | p75 | Std Dev | max |
|---|---|---|---|---|---|---|---|---|
| Agricultural sciences | Cites | 2684 | 8.3 | 1.0 | 2.0 | 5.0 | 40.9 | 973.0 |
| | $M_i$ | | 177.3 | 29.0 | 110.0 | 271.0 | 186.0 | 727.0 |
| | $M_j$ | | 2111.0 | 240.1 | 801.4 | 2371.6 | 3401.4 | 15245.4 |
| | $d_{ij}$ | | 366.6 | 155.1 | 320.9 | 546.8 | 258.0 | 1084.5 |
| Engineering and technology | Cites | 4059 | 26.3 | 1.0 | 3.0 | 8.0 | 256.3 | 8302.0 |
| | $M_i$ | | 930.5 | 62.0 | 386.0 | 990.0 | 1379.0 | 5321.0 |
| | $M_j$ | | 3343.4 | 288.1 | 1337.1 | 3949.3 | 5227.9 | 22115.6 |
| | $d_{ij}$ | | 361.9 | 141.7 | 309.4 | 548.8 | 264.2 | 1075.0 |
| Humanities | Cites | 492 | 3.6 | 1.0 | 1.0 | 3.0 | 10.4 | 156.0 |
| | $M_i$ | | 122.8 | 25.0 | 54.0 | 165.0 | 145.1 | 546.0 |
| | $M_j$ | | 2741.2 | 622.1 | 1140.1 | 3638.2 | 3471.8 | 12493.6 |
| | $d_{ij}$ | | 320.2 | 116.1 | 246.5 | 491.3 | 259.8 | 1021.8 |
| Medical and health sciences | Cites | 7103 | 30.2 | 1.0 | 2.0 | 7.0 | 380.1 | 20385.0 |
| | $M_i$ | | 2108.2 | 63.0 | 471.0 | 1864.0 | 4066.8 | 17259.0 |
| | $M_j$ | | 4780.4 | 138.9 | 1103.8 | 4560.4 | 9382.5 | 41008.7 |
| | $d_{ij}$ | | 361.5 | 149.9 | 303.1 | 537.0 | 261.2 | 1119.0 |
| Natural sciences | Cites | 6593 | 41.5 | 1.0 | 3.0 | 10.0 | 493.3 | 25797.0 |
| | $M_i$ | | 1712.2 | 94.0 | 538.0 | 1902.0 | 2847.7 | 13337.0 |
| | $M_j$ | | 4173.3 | 197.3 | 978.9 | 4725.8 | 7195.3 | 34110.7 |
| | $d_{ij}$ | | 372.7 | 154.5 | 330.3 | 556.0 | 261.7 | 1084.5 |
| Social sciences | Cites | 1912 | 9.9 | 1.0 | 2.0 | 4.0 | 77.8 | 2342.0 |
| | $M_i$ | | 365.6 | 45.0 | 143.0 | 380.0 | 574.4 | 2049.0 |
| | $M_j$ | | 2548.5 | 256.9 | 1051.6 | 2789.8 | 3958.1 | 15603.9 |
| | $d_{ij}$ | | 358.3 | 142.1 | 294.4 | 540.2 | 267.2 | 1083.8 |

Table 2 shows the estimates of the coefficients of the gravitational model calculated by means of OLS. The R-squared values are always lower than 0.6; the lowest values are recorded in Humanities (0.398) and in Social sciences (0.471).

Our next focus is on the variable of interest $d_{ij}$. After demonstrating at an aggregate level that distance still matters in scholarly knowledge flows in science (Abramo, D'Angelo, & Di Costa, 2019), the findings confirm that the same phenomenon is also present at a lower level of aggregation, but with different intensities. For all six DAs considered there is a clear effect of geographic proximity on knowledge flows, with values of γ all negative and statistically significant: a percentage increase of 1% in distance corresponds to a decrease in the citations exchanged that varies in the range of 0.3%-0.5%, in absolute value. In detail, the most significant reductions on averages of citations exchanged are observed in Natural sciences (-0.505) and Engineering and technology (-0.497), followed by the closely grouped threesome of Medical and health sciences, Social sciences and Agricultural sciences (respectively -0.427, -0.409, -0.392), and finally at a distance, Humanities (-0.277).

In the national context, the geographic proximity effect on citations between territories results as present, but differentiated by DAs: more contained within the Humanities and Social sciences, more significant in the Sciences, in particular in Natural sciences and Engineering and technology. These data, corroborated by the descriptive statistics of Table 1, suggest the hypothesis that Humanities and Social sciences are characterized by the limitation of geographic influence to a small area, probably reflecting the national specificity of the research topics covered.

The same applies for the variables $M_i$ and $M_j$, whose coefficients are all positive and statistically significant: for these, a percentage increase of 1% corresponds to an increase in citations exchanged that varies in the range 0.1%÷0.5%, with the minimums always in Humanities (0.114 for $M_i$, 0.145 for $M_j$) and in Social sciences (0.296 for $M_i$, 0.279 for $M_j$).

*Table 2: OLS regression outcome at the disciplinary area level for the national context*

|  | Obs | $M_i$ | | $M_j$ | | $d_{ij}$ | | Const | | $R^2$ |
|---|---|---|---|---|---|---|---|---|---|---|
| Agricultural sciences | 2684 | 0.323 | *** | 0.251 | *** | -0.392 | *** | 0.066 | ns | 0.445 |
| Engineering and technology | 4059 | 0.385 | *** | 0.390 | *** | -0.497 | *** | -0.765 | *** | 0.533 |
| Humanities | 492 | 0.114 | *** | 0.145 | *** | -0.277 | *** | 0.536 | *** | 0.398 |
| Medical and health sciences | 7103 | 0.413 | *** | 0.424 | *** | -0.427 | *** | -1.634 | *** | 0.532 |
| Natural sciences | 6593 | 0.438 | *** | 0.439 | *** | -0.505 | *** | -1.325 | *** | 0.564 |
| Social sciences | 1912 | 0.296 | *** | 0.279 | *** | -0.409 | *** | -0.135 | ns | 0.471 |

*Y= cites*
*Significance level: \*\*\* 0.01, \*\* 0.05, \* 0.1*

### 3.1.2 The international context

Table 3 shows the estimates of the coefficients at DA level for the analysis of the continental European (EUR) context. The R-squared values are always greater than 0.5, except for Humanities (0.337). The effect of geographic proximity on knowledge flows is evident in all six DAs, with values of $\gamma$ all negative and statistically significant. A 1% increase in distance corresponds to a decrease in citations of around 0.2%, with the lowest values recorded in Social sciences (-0.119), Humanities (-0.151) and Engineering and technology (-0.193). In contrast, given increasing distance, the most significant reductions in citations are observed in Natural sciences (-0.261), Agricultural sciences (-0.250), Medical and health sciences (-0.215).

As seen previously in the national case, the values of the coefficients of the variables $M_i$ and $M_j$, all positive and statistically significant, are almost aligned; however, these always reveal Humanities as the DA with the lowest coefficients (0.265 for $M_i$, 0.341 for $M_j$).

*Table 3: OLS regression outcome at the disciplinary area level for the European context*

|  | Obs | $M_i$ | | $M_j$ | | $d_{ij}$ | | Const | | $R^2$ |
|---|---|---|---|---|---|---|---|---|---|---|
| Agricultural sciences | 2230 | 0.577 | *** | 0.570 | *** | -0.250 | *** | -4.771 | *** | 0.572 |
| Engineering and technology | 3429 | 0.667 | *** | 0.736 | *** | -0.193 | *** | -7.675 | *** | 0.709 |
| Humanities | 600 | 0.265 | *** | 0.341 | *** | -0.151 | ** | -2.705 | *** | 0.337 |
| Medical and health sciences | 5314 | 0.742 | *** | 0.786 | *** | -0.215 | *** | -8.376 | *** | 0.750 |
| Natural sciences | 5132 | 0.713 | *** | 0.777 | *** | -0.261 | *** | -7.794 | *** | 0.747 |
| Social sciences | 1991 | 0.600 | *** | 0.657 | *** | -0.119 | *** | -6.918 | *** | 0.615 |

Finally, we carry out the same analysis out for the extra-EUR context (Table 4). The results show R-squared varying in the range 0.4÷0.7, with the values higher than 0.6 for all DAs except Humanities (0.4).

For four out of the six DAs considered, the geographic effect on knowledge flows would seem attested by statistically significant values and positive in sign. For the remaining two DAs (Humanities and Medical and health sciences) the geographic proximity effect is not manifested, since their p-values are not statistically significant. Taking this evidence as a whole, the geographic effect clearly disappears beyond a "threshold distance", meaning that the phenomenon would be confined to the national and continental scale.

We can hypothesize that as distances increase, the contact mediated by information and communications technologies prevails over purely "personal" relationships. In the four DAs where the values of the γ coefficient are statistically significant, these are in any case all lower than 0.12, in absolute value. Still, although limited in value, it remains to be understood why there would be a positive sign on the γ coefficient of dij for the four DAs, in considering the intercontinental citation flows.

*Table 4: OLS regression outcome at the disciplinary area level for the extra-European context*

|  | Obs | $M_i$ | | $M_j$ | | $d_{ij}$ | | Const | | $R^2$ |
|---|---|---|---|---|---|---|---|---|---|---|
| Agricultural sciences | 2377 | 0.584 | *** | 0.538 | *** | 0.063 | ** | -6.949 | *** | 0.615 |
| Engineering and technology | 3636 | 0.687 | *** | 0.743 | *** | 0.063 | *** | -9.910 | *** | 0.740 |
| Humanities | 363 | 0.259 | *** | 0.375 | *** | -0.081 | ns | -3.651 | *** | 0.398 |
| Medical and health sciences | 5642 | 0.662 | *** | 0.752 | *** | 0.010 | ns | -9.487 | *** | 0.693 |
| Natural sciences | 5586 | 0.691 | *** | 0.789 | *** | 0.097 | *** | -10.813 | *** | 0.738 |
| Social sciences | 1738 | 0.549 | *** | 0.588 | *** | 0.119 | *** | -8.182 | *** | 0.610 |

### 3.2 Analysis at the level of subject category

We can now replicate the analysis seen at the DA level, but at the SC level. This is a critical analysis, given that the DAs aggregate SCs, which in addition to varying contents, also have different characteristics in terms of publication intensity and citability.

Table 5 shows, as an example, the results for the 11 SCs belonging to the Agricultural sciences DA.

The results of the OLS regression show that the R-squared values vary in the range 0.4÷0.5 (national case), 0.2÷0.5 (EUR), and 0.3÷0.6 (extra-EUR). Food Science & Technology has the highest values in both EUR and extra-EUR contexts. In the analysis at national scale, the coefficients relating to distances are all statistically significant and negative. In the European context this is true for 7 SCs (with the exception of Agricultural Engineering, Agricultural Economics & Policy, Fisheries, Horticulture), and in the extra-European context for just two (Agriculture, dairy & animal science; Food Science & Technology).

We can conclude that the presence of the geographic proximity effect is also confirmed at the SC level, certainly for the citation flows on a national scale and to a less extent in the European context. In the case of extra-EUR the geographic proximity effect is almost always not significant and possibly confined to a limited number of SCs.

This is confirmed by the data of Table 6, which for each DA presents the descriptive statistics for the distribution of the γ coefficient for the variable $d_{ij}$. The maximum and minimum values thus refer to what is observed for the SCs of the given DA.

At national scale, the coefficients are always all negative: geographic distance between municipalities has a negative impact on the citation flows in all SCs. Instead, the "Max" column evidences some positive values, i.e. the presence of at least one SC where geographic proximity reduces the citation flows, both in the continental and inter-continental analyses. At EUR scale, this is recorded in six SCs belonging to four different DAs: Engineering and technology (SC of Engineering, Chemical), Humanities (SCs of Art and History), Social Sciences (Criminology & penology and Ergonomics), Medical and Health Sciences (Nursing). On the Extra-EUR scale, positive values are recorded in 26 SCs belonging to five different DAs: more precisely in nine SCs of Natural sciences, eight of Medical and health sciences, six of Engineering and technology, two of Social Sciences and one of Agricultural sciences. The "Min" column evidences one case only (Humanities), on the EUR scale, showing all positive values for each relevant SC.

Comparing columns 2 and 5 of Table 6, we observe that the minimum γ coefficients in the national context are always higher, in absolute value, than those recorded in the EUR context. We can conclude that the geographic bias, where significant, is always greater for national flows than for continental ones.

It is also interesting to observe the trend in the value of standard deviation within each DA, i.e. the dispersion of data around the average. In the national context, the highest value for dispersion of data is observed in Engineering and technology and Natural sciences, while the lowest corresponds to Humanities; in the other DAs the values are almost equal. In the EUR context, Social sciences presents the maximum value of standard deviation, Agricultural sciences the minimum. In the extra-EUR context, the highest and lowest values occur respectively in Social sciences, and Humanities.

*Table 5: OLS regression outcome for the subject categories of the Agricultural sciences disciplinary area*

| | Italy | | | | EUR | | | | Extra-EUR | | | |
|---|---|---|---|---|---|---|---|---|---|---|---|---|
| SubCat | Obs | $d_{ij}$ coeff. | | $R^2$ | Obs | $d_{ij}$ coeff. | | $R^2$ | Obs | $d_{ij}$ coeff. | | $R^2$ |
| Agriculture, dairy & animal science | 451 | -0.348 *** | 0.027 | 0.485 | 580 | -0.115 * | 0.064 | 0.271 | 534 | -0.129 ** | 0.057 | 0.351 |
| Agricultural Engineering | 327 | -0.255 *** | 0.025 | 0.440 | 423 | 0.005 ns | 0.075 | 0.194 | 514 | 0.061 ns | 0.054 | 0.383 |
| Agricultural Economics & Policy | 54 | -0.151 *** | 0.037 | 0.402 | 92 | -0.128 ns | 0.124 | 0.298 | 84 | -0.120 ns | 0.178 | 0.325 |
| Agriculture, Multidisciplinary | 668 | -0.311 *** | 0.021 | 0.439 | 750 | -0.104 * | 0.060 | 0.295 | 771 | 0.007 ns | 0.040 | 0.340 |
| Agronomy | 637 | -0.302 *** | 0.024 | 0.387 | 710 | -0.124 ** | 0.057 | 0.366 | 716 | -0.049 ns | 0.043 | 0.451 |
| Fisheries | 241 | -0.216 *** | 0.028 | 0.407 | 373 | -0.006 ns | 0.082 | 0.210 | 357 | -0.051 ns | 0.055 | 0.305 |
| Food Science & Technology | 1701 | -0.375 *** | 0.019 | 0.436 | 1514 | -0.167 *** | 0.041 | 0.513 | 1588 | 0.053 * | 0.031 | 0.567 |
| Forestry | 371 | -0.299 *** | 0.022 | 0.438 | 509 | -0.118 * | 0.065 | 0.412 | 387 | -0.066 ns | 0.074 | 0.439 |
| Horticulture | 510 | -0.297 *** | 0.025 | 0.406 | 520 | -0.016 ns | 0.061 | 0.264 | 535 | 0.001 ns | 0.046 | 0.367 |
| Soil Science | 349 | -0.295 *** | 0.026 | 0.448 | 460 | -0.229 *** | 0.084 | 0.345 | 440 | 0.073 ns | 0.060 | 0.448 |
| Veterinary Sciences | 591 | -0.373 *** | 0.023 | 0.516 | 777 | -0.134 ** | 0.060 | 0.479 | 734 | -0.034 ns | 0.043 | 0.547 |
| Total Area | 2684 | -0.392 *** | 0.016 | 0.445 | 2230 | -0.250 *** | 0.036 | 0.572 | 2377 | 0.063 ** | 0.027 | 0.615 |

*Significance level: \*\*\* 0.01, \*\* 0.05, \* 0.1. In brackets, robust standard error for the $d_{ij}$ γ coefficient*

*Table 6: Descriptive statistics for the distribution of the $d_{ij}$ γ coefficient for the SCs of each disciplinary area\**

|  | Italy | | | EUR | | | Extra-EUR | | |
|---|---|---|---|---|---|---|---|---|---|
| Area_OECD | Min | Max | St.dev. | Min | Max | St.dev. | Min | Max | St.dev. |
| Agricultural sciences | -0.375 | -0.151 | 0.063 | -0.229 | -0.104 | 0.040 | -0.129 | 0.053 | 0.091 |
| Engineering and technology | -0.495 | -0.163 | 0.077 | -0.321 | 0.092 | 0.073 | -0.192 | 0.156 | 0.128 |
| Humanities | -0.256 | -0.119 | 0.043 | 0.170 | 0.372 | 0.101 | -0.268 | -0.164 | 0.052 |
| Medical and health sciences | -0.421 | -0.149 | 0.065 | -0.334 | 0.238 | 0.092 | -0.215 | 0.267 | 0.139 |
| Natural sciences | -0.508 | -0.148 | 0.076 | -0.434 | -0.092 | 0.073 | -0.300 | 0.195 | 0.146 |
| Social sciences | -0.349 | -0.099 | 0.066 | -0.278 | 0.365 | 0.184 | -0.891 | 0.172 | 0.315 |

*\* Limited to the SCs with more than 30 observations and with significant $d_{ij}$ γ coefficient*

Table 7 shows the counts concerning distribution of the coefficient γ of the $d_{ij}$ variable in the three territorial contexts analyzed. In the analysis of flows between national municipalities, of the 215 cases where the γ coefficient is significant, it is also systematically negative; in the analysis of continental flows, the SCs with significant γ coefficient drop to 125, and in six of these the sign is positive, indicating that rather than distance limiting flows, they are encouraged. Finally, in the extra-EUR context, the number of SCs where the OLS model returns significant γ coefficients drops further, to 59, and in 26 of these the γ coefficient is not negative.

*Table 7: Counting data for $d_{ij}$ γ coefficient by disciplinary area and subject category*

|  | Italy | | EUR | | Extra-EUR | |
|---|---|---|---|---|---|---|
| Area | No. of SCs with γ significant | Of which with negative sign | No. of SCs with g significant | Of which with negative sign | No. of SCs with γ significant | Of which with negative sign |
| Agricultural sciences | 11 | 11 | 7 | 7 | 2 | 1 |
| Engineering and technology | 42 | 42 | 24 | 23 | 10 | 4 |
| Humanities | 6 | 6 | 2 | 0 | 2 | 2 |
| Medical and health sciences | 56 | 56 | 33 | 32 | 17 | 9 |
| Natural sciences | 63 | 63 | 48 | 48 | 20 | 11 |
| Social sciences | 37 | 37 | 11 | 9 | 8 | 6 |
| Total | 215 | 215 | 125 | 119 | 59 | 33 |

## 4. Conclusions

The current work continues from and deepens a previous study by the authors (Abramo, D'Angelo & Di Costa; 2019), concerning the influence of geographic distance on the knowledge flows from producers of new knowledge (articles cited) to the users (articles citing). In this case the specific aim is to study and compare the knowledge diffusion across scientific fields, for determination of if and how the effects of geographic distance vary between SCs. The study is based on the same methodological assumptions of the previous paper: using a gravitational model estimated by ordinary least squares, we have now analyzed the effect of geographic proximity in each DA and SC, at three different geographic scales.

On a national scale, the results show that as the distance between territories increases, controlling for their mass, there is a decrease in the number of citations exchanged in all SCs investigated, a finding in line with previous literature.

However the proximity effect is more evident in Natural sciences, Engineering and technology, and less in Humanities. The same occurs on a European scale, but with less noticeable decreases than observed at national scale: in relative terms, the reductions in knowledge flows with distance are more appreciable in the DAs of Natural sciences and Agricultural sciences, less in Social sciences and Humanities. In the analysis carried out at extra-European scale, the geographic effect seems to disappear; instead there is even a positive relationship between quotations exchanged and distance, in Engineering and technology and in Natural sciences, while the inverse relationship holds only in Humanities and Agricultural sciences.

Therefore, what we previously observed at the aggregate level is confirmed at SC level, concerning the presence of a "threshold" effect beyond which the geographic distance effect disappears and the intensity of the citation flows becomes insensitive to the spatial factor. It is evident that at the intercontinental scale, ICT-mediated communications have been progressively replacing face-to-face contacts, and so the effects of geographic proximity have waned. However, such results could also be linked to the geographic context analyzed. This type of analysis is inevitably country specific, as is the very concept of "intercontinental": each country has its own specific place in the world and what is evident for the flows generated by Italian scientific production might not be so in other cases, for example the New Zealand one. It follows that a fundamental aspect of these types of analysis is that they must necessarily consider geographic scale as a fundamental factor. This would be true for Italy, in particular, whose scientific production generates extra-EUR citation flows that compose almost 50% of the total international ones.

This work also evidences that geographic bias tends to be differentiated between SCs, by virtue of their intrinsic characteristics. Humanities and Social sciences have a smaller area of influence, a smaller average range of citation flows; at the same time, the decay of citation flows with geographic distance is lower than in other DAs, particularly compared to the Sciences. This could be due to the peculiarity of the research topics addressed, more country specific for Humanities and Social sciences, and therefore with more localized spillovers, but also with lower citability of the works, as well as lower incidence of self-citations for these DAs compared to the SCs of Sciences. All these determinants can be the object of further study, continuing from the current work.

Still on the subject of future developments, from a methodological point of view it could be useful to work on the specification of the model, for example by integrating a series of latent variables associated with the so-called "social proximity factors" (links between mentors and students, belonging to the same scientific school within a field; asymmetry in citation processes in favour of papers published in prestigious journals, or by prestigious scientists, ...). In fact citations reflect not only the attribution of scientific credit, but can also be dictated by reasons of a "social" nature, and this could generate a bias in favor of the geographic factor. Finally, it would certainly be interesting to include the time variable in the analytical model, with the aim of verifying the variation of the effect of geographic distance as a function of time, with the relevant implications concerning citation time windows.

# APPENDIX

*Table A1: Descriptive statistics for the European context analysis by disciplinary area*

| Disciplinary area | Var | No. | Mean | p25 | p50 | p75 | Std dev | Cv | Max |
|---|---|---|---|---|---|---|---|---|---|
| Agricultural sciences | Cites | 2230 | 14.6 | 1.0 | 4.0 | 12.0 | 35.7 | 2.5 | 598 |
| | *Mi* | 2230 | 123.6 | 16.0 | 63.0 | 200.0 | 157.4 | 1.3 | 727 |
| | *Mj* | 2230 | 58296.3 | 15861.5 | 33956.0 | 61393.7 | 63528.8 | 1.1 | 236707 |
| | *dij* | 2230 | 1291.8 | 824.6 | 1261.7 | 1681.7 | 564.3 | 0.4 | 3789 |
| Engineering and technology | Cites | 3429 | 38.0 | 2.0 | 5.0 | 21.0 | 134.4 | 3.5 | 2762 |
| | *Mi* | 3429 | 515.4 | 21.0 | 94.0 | 515.0 | 1008.9 | 2.0 | 5321 |
| | *Mj* | 3429 | 90470.8 | 18591.5 | 51344.3 | 99977.1 | 100335.3 | 1.1 | 373193 |
| | *dij* | 3429 | 1310.0 | 826.6 | 1285.5 | 1704.6 | 576.3 | 0.4 | 3561 |
| Humanities | Cites | 600 | 4.0 | 1.0 | 2.0 | 4.0 | 7.4 | 1.8 | 82 |
| | *Mi* | 600 | 106.5 | 21.0 | 51.0 | 141.0 | 134.0 | 1.3 | 546 |
| | *Mj* | 600 | 67110.3 | 21804.2 | 38725.8 | 103078.5 | 70262.5 | 1.0 | 233547 |
| | *dij* | 600 | 1265.8 | 843.0 | 1237.9 | 1649.7 | 518.9 | 0.4 | 3301 |
| Medical and health sciences | Cites | 5314 | 82.5 | 2.0 | 5.0 | 22.0 | 494.1 | 6.0 | 14653 |
| | *Mi* | 5314 | 778.3 | 17.0 | 71.0 | 471.0 | 2309.8 | 3.0 | 17259 |
| | *Mj* | 5314 | 123703.6 | 25584.9 | 67422.3 | 138529.0 | 144141.1 | 1.2 | 520222 |
| | *dij* | 5314 | 1290.2 | 812.1 | 1252.1 | 1676.6 | 576.9 | 0.4 | 3837 |
| Natural sciences | Cites | 5132 | 89.3 | 2.0 | 6.0 | 30.0 | 457.0 | 5.1 | 11820 |
| | *Mi* | 5132 | 789.1 | 14.0 | 128.0 | 685.0 | 1861.9 | 2.4 | 13337 |
| | *Mj* | 5132 | 129379.0 | 22010.3 | 73521.2 | 138737.6 | 149879.4 | 1.2 | 548509 |
| | *dij* | 5132 | 1310.0 | 817.1 | 1283.4 | 1699.2 | 591.3 | 0.5 | 3837 |
| Social sciences | Cites | 1991 | 19.8 | 1.0 | 4.0 | 11.0 | 74.0 | 3.7 | 1416 |
| | *Mi* | 1991 | 217.2 | 21.0 | 65.0 | 160.0 | 425.8 | 2.0 | 2049 |
| | *Mj* | 1991 | 66133.7 | 22423.8 | 36747.4 | 61651.8 | 76170.6 | 1.2 | 297825 |
| | *dij* | 1991 | 1322.7 | 842.4 | 1290.5 | 1712.1 | 577.6 | 0.4 | 3837 |

*Table A2: Descriptive statistics for the extra-European context analysis by disciplinary area*

| Disciplinary area | Var | No. | Mean | p25 | p50 | p75 | Std dev | Cv | Max |
|---|---|---|---|---|---|---|---|---|---|
| Agricultural sciences | Cites | 2377 | 16.1 | 1.0 | 3.0 | 10.0 | 57.7 | 3.6 | 1630 |
|  | $M_i$ | 2377 | 151.4 | 19.0 | 83.0 | 230.0 | 182.4 | 1.2 | 727 |
|  | $M_j$ | 2377 | 116920.5 | 7110.0 | 24446.4 | 112165.8 | 217897.1 | 1.9 | 869055 |
|  | $d_{ij}$ | 2377 | 7773.8 | 4980.4 | 8058.5 | 9759.9 | 3929.8 | 0.5 | 18898 |
| Engineering and technology | Cites | 3636 | 60.5 | 1.0 | 4.0 | 18.0 | 335.8 | 5.6 | 8061 |
|  | $M_i$ | 3636 | 639.6 | 33.0 | 135.0 | 626.0 | 1153.4 | 1.8 | 5321 |
|  | $M_j$ | 3636 | 207736.3 | 11439.9 | 40538.3 | 207613.5 | 380346.0 | 1.8 | 1259611 |
|  | $d_{ij}$ | 3636 | 7596.2 | 4416.7 | 8028.9 | 9707.4 | 3822.8 | 0.5 | 18898 |
| Humanities | Cites | 363 | 4.7 | 1.0 | 2.0 | 4.0 | 11.6 | 2.5 | 148 |
|  | $M_i$ | 363 | 121.7 | 25.0 | 54.0 | 165.0 | 147.5 | 1.2 | 546 |
|  | $M_j$ | 363 | 197999.2 | 22226.7 | 57115.2 | 129472.5 | 283426.1 | 1.4 | 850975 |
|  | $d_{ij}$ | 363 | 8287.5 | 6482.2 | 8039.4 | 9797.3 | 3881.0 | 0.5 | 18890 |
| Medical and health sciences | Cites | 5642 | 111.9 | 1.0 | 3.0 | 15.0 | 1164.9 | 10.4 | 55437 |
|  | $M_i$ | 5642 | 1319.0 | 23.0 | 108.0 | 1158.0 | 3229.7 | 2.4 | 17259 |
|  | $M_j$ | 5642 | 233912.5 | 6336.4 | 37095.2 | 187383.5 | 513271.7 | 2.2 | 2172464 |
|  | $d_{ij}$ | 5642 | 7664.0 | 4652.5 | 7948.9 | 9672.7 | 3761.9 | 0.5 | 18916 |
| Natural sciences | Cites | 5586 | 120.9 | 1.0 | 4.0 | 21.0 | 1001.3 | 8.3 | 40171 |
|  | $M_i$ | 5586 | 1236.2 | 29.0 | 241.0 | 1183.0 | 2466.4 | 2.0 | 13337 |
|  | $M_j$ | 5586 | 240076.9 | 5925.4 | 46635.5 | 230767.6 | 478789.8 | 2.0 | 1903310 |
|  | $d_{ij}$ | 5586 | 7621.5 | 4460.4 | 7956.1 | 9710.8 | 3862.1 | 0.5 | 18898 |
| Social sciences | Cites | 1738 | 22.8 | 1.0 | 3.0 | 9.0 | 128.4 | 5.6 | 3029 |
|  | $M_i$ | 1738 | 293.6 | 25.0 | 91.0 | 193.0 | 527.4 | 1.8 | 2049 |
|  | $M_j$ | 1738 | 139492.8 | 11603.4 | 28771.8 | 143990.6 | 274480.5 | 2.0 | 1100241 |
|  | $d_{ij}$ | 1738 | 8052.3 | 5564.0 | 8147.0 | 9794.2 | 3877.4 | 0.5 | 18890 |